\pdfoutput=1
\documentclass[aps,prd,showpacs,nofootinbib,10pt,superscriptaddress,twocolumn,amsmath,amssymb,floatfix,showkeys]{revtex4-1}
\usepackage{color,amsmath,amssymb,graphicx,latexsym,lineno}
\usepackage{threeparttable,multirow,txfonts,booktabs}
\usepackage{accents,slashed,ulem,subfig}
\usepackage[para]{footmisc}
\definecolor{blue0}{rgb}{0,0,0.6}
\usepackage[colorlinks,linkcolor=blue0,anchorcolor=blue0,citecolor=blue0,urlcolor=blue0]{hyperref}

\newlength{\dhatheight}
\newcommand{\doublehat}[1]{%
    \settoheight{\dhatheight}{\ensuremath{\hat{#1}}}%
    \addtolength{\dhatheight}{-0.15ex}%
    \hat{\vphantom{\rule{1pt}{\dhatheight}}%
    \smash{\hat{#1}}}}
\graphicspath{{figs/}}

\begin{document}

\title{Constraints on Axion-like Particles from Observations of Mrk 421 using the ${\rm CL_s}$ Method}

\author{Lin-Qing Gao$^*$}
\affiliation{School of Nuclear Science and Technology, University of South China, Hengyang 421001, China}

\author{Xiao-Jun Bi$^\dag$}
\affiliation{Key Laboratory of Particle Astrophysics, Institute of High Energy Physics,
Chinese Academy of Sciences, Beijing 100049, China}
\affiliation{School of Physical Sciences, University of Chinese Academy of Sciences, Beijing, China}

\author{Jun-Guang Guo$^\ddag$}
\affiliation{School of Physics and Electronic Engineering, Jiaying University, Meizhou 514015, China}

\author{Wenbin Lin$^\S$}
\affiliation{School of Mathematics and Physics, University of South China,
Hengyang, 421001, China}

\author{Peng-Fei Yin$^\P$}
\affiliation{Key Laboratory of Particle Astrophysics, Institute of High Energy Physics,
Chinese Academy of Sciences, Beijing 100049, China}

%\date{\today}

\begin{abstract}

Axion-like particles (ALPs) may undergo mixing with photons in the presence of astrophysical magnetic fields, leading to alterations in the observed high energy $\gamma$-ray spectra. In this study, we investigate the ALP-photon oscillation effect using the spectra of the blazar Mrk 421 over 15 observation periods measured by Major Atmospheric Gamma Imaging Cherenkov Telescopes (MAGIC) and Fermi Large Area Telescope (Fermi-LAT). Compared with previous studies, we generate mock data under the ALP hypothesis and employ the ${\rm CL_s}$ method to set constraints on the ALP parameters.  This method is widely utilized in high energy experiments and avoids the exclusion of specific parameter regions where distinguishing between the null and ALP hypotheses is challenging. We find that the ALP-photon coupling $g_{a\gamma}$ is constrained to be smaller than $\sim 2\times10^{-11}$ GeV$^{-1}$ for ALP masses ranging from $10^{-9}$ eV to $10^{-7}$ eV at the 95\%  confidence level. We also present the constraints derived from the TS distribution under the null hypothesis, which is commonly utilized in previous astrophysical ALP studies. Our results reveal that the combined constraints of all the periods obtained from both methods are consistent. However, the ${\rm CL_s}$  method remains effective in cases where the latter method fails to provide constraints for specific observation periods.

\end{abstract}

\maketitle

\section{introduction}

Axions are light pseudo scalar particles predicated by the Peccei-Quinn mechanism, which can elegantly solve the strong CP problem in the quantum chromodynamics \cite{Peccei:1977ur,Peccei:2006as,Weinberg:1977ma, Wilczek:1977pj}. More generally, various extensions of the standard model, including string theory \cite{Svrcek:2006yi, Arvanitaki:2009fg, Marsh:2015xka}, predict the so-called axion-like particles (ALPs), which are also light pseudo scalars but do not necessarily solve the strong CP problem. The mass and coupling of the quantum chromodynamics axion are related, but they are not necessarily related for ALPs \cite{Jaeckel:2010ni}.
This means that ALPs have a broader parameter space and rich phenomenology to be explored.

The interactions between the ALP and standard model particles are known to be very weak. One prominent interaction is the vertex between one ALP and two photons, which could lead to ALP-photon conversion in the presence of external magnetic or electric fields \cite{Raffelt:1987im}. Experimental searches for ALP typically fall into several categories, including light-shining-through-the-wall experiments (such as ALPS \cite{Ehret:2010mh}), solar ALP helioscopes (such as CAST \cite{CAST:2017uph}), underground detectors (such as CDEX \cite{CDEX:2018lau} and XENON1T \cite{XENON:2018voc}), and haloscopes (for the ALP dark matter, such as ADMX \cite{ADMX:2006kgb}). 

The oscillation between the ALP and photons can exert an influence on the photon spectra of astrophysical sources and has garnered significant research interest \cite{DeAngelis:2007dqd, Hooper:2007bq, Simet:2007sa, Mirizzi:2007hr, Belikov:2010ma, DeAngelis:2011id, Horns:2012kw, HESS:2013udx, Meyer:2013pny, Tavecchio:2014yoa, Meyer:2014epa, Meyer:2014gta, Fermi-LAT:2016nkz, Meyer:2016wrm, Berenji:2016jji, Galanti:2018upl, Galanti:2018myb, Zhang:2018wpc, Liang:2018mqm, Bi:2020ths, Guo:2020kiq, Li:2020pcn, Li:2021gxs, Cheng:2020bhr, Liang:2020roo, Xia:2018xbt}. As photons propagate, they encounter a series of magnetic field environments along the line of sight, leading to continuous ALP-photon oscillation. The oscillation depends on the energy of the photons and leads to irregularities in the observed spectrum. For extragalactic sources, emitted photons may convert to ALPs efficiently in the ambient magnetic field of the source, then travel unimpeded in interstellar space, and finally ALPs can convert back to photons in the Galactic magnetic field. This process reduces the absorption effect of the high-energy photons from extragalactic sources induced by the extragalactic background light (EBL). 
With the advancements in high-energy astrophysical experiments, 
such as Fermi-LAT \cite{Fermi-LAT:2009ihh}, MAGIC \cite{MAGIC:2014zas}, HAWC \cite{HAWC:2020hrt}, Argo-YBJ \cite{ARGO-YBJ:2015qiq}, and LHAASO \cite{LHAASO:2019qtb}, precise $\gamma$-ray spectra of many extragalactic sources have been measured. These observations provide an opportunity to investigate the effects of ALP-photon oscillation.

Blazars are a type of active galactic nucleus that contains a supermassive black hole surrounded by an accretion disk and emits two jets perpendicular to the disk. Blazars have emerged as particularly promising targets for investigating the ALP-photon oscillation effect due to their brightness and dominant presence in extragalactic $\gamma$-ray emissions.
Numerous studies have been performed to set constraints on the ALP parameter space in the literature \cite{Belikov:2010ma,DeAngelis:2011id,Horns:2012kw,HESS:2013udx,Meyer:2013pny,Meyer:2014epa,Meyer:2014gta,Galanti:2018myb,Galanti:2018upl,Zhang:2018wpc,Guo:2020kiq,Li:2020pcn,Li:2021gxs}.

In order to set constraints on the ALP parameters, one can define the test statistic (TS) as the logarithmic ratio of the best-fit likelihoods under the null hypothesis without ALP and the alternative hypothesis with ALP. However, due to the nonlinear relation between the ALP parameters and the modifications of the photon spectra, the commonly used Wilks' theorem \cite{wilk} is not appropriately applied here \cite{Fermi-LAT:2016nkz}. Consequently, the TS values would not follow the ordinary $\chi^2$ distribution. Therefore, Monte Carlo simulations are necessary to obtain the realistic TS distribution.
In principle, the mock data is required to be generated for each parameter point under the alternative hypothesis \cite{HESS:2013udx, Buehler:2020qsn, Cheng:2020bhr}.
However, for the sake of saving time, many studies do not directly perform  simulations for the alternative hypothesis, and instead adopt the TS distribution under the null hypothesis in the statistical analysis (referred to as the simplified method hereafter) \cite{Fermi-LAT:2016nkz, Guo:2020kiq, Li:2020pcn, Li:2021gxs, Xia:2018xbt, Liang:2018mqm, Xia:2019yud}.

Note that when the signal is not sufficiently significant, the TS distributions of the alternative and null hypotheses are not well separated. This can lead to overly optimistic limits when the data fluctuates from the prediction of the null hypothesis. In order to address this issue, we simulate the mock data for each ALP parameter point under the alternative hypothesis and utilize the ${\rm CL_s}$ method \cite{Junk:1999kv,Read:2002hq_cls,Lista:2016chp}, which is widely employed in high energy experiments, in the statistical analysis. This method avoids the exclusion of certain parameter regions where distinguishing between the alternative and null hypotheses is challenging., thus providing more conservative constraints. For comparison, we also present the constraints obtained from the simplified method. 

In this study, we analyze the very high energy $\gamma$-ray spectra of Mrk 421 observed by MAGIC and Fermi-LAT during 2013 and 2014, which include a total of 15 observation periods \cite{MAGIC:2019ozu}. Mrk 421 is the brightest extragalactic source with a redshift of $z=0.031$ and is classified as a high-frequency-peaked BL Lac object. Due to its flaring nature, each of the 15 observation periods can be treated as independent data sets, enabling us to analyze them separately. Furthermore, combining the results from these periods generally yield a more stringent constraint on the ALP parameters.

This paper is structured as follows. In Sec. \ref{sec:ALPs and magnetic}, we give a brief introduction to the ALP-photon oscillation effect and the astrophysical magnetic fields considered in the line of sight. In Sec. \ref{sec:method}, we show the methods about the spectrum fitting and statistical analysis on the ALP parameters .
Next, in Sec. \ref{sec:result}, we give constraints on the ALP parameter plane. Lastly, we give a summary in Sec. \ref{sec:summary}.

\section{ALP-photon oscillation in the magnetic field}\label{sec:ALPs and magnetic}
\subsection{ALP-photon oscillation}
The Lagrangian describing the interaction between the ALP and photons is
\begin{equation}
\mathcal{L}_{a\gamma} = -\frac{1}{4} g_{a\gamma} a F_{\mu \nu} \Tilde{F}^{\mu \nu} = g_{a\gamma} a \boldsymbol{E} \cdot \boldsymbol{B},
\end{equation}
where $g_{a\gamma}$ denotes the coupling parameter, $F$ denotes the electromagnetic field-strength tensor, $a$ denotes the ALP field, $\boldsymbol{E}$ represents the electric field, and $\boldsymbol{B}$ represents the magnetic field. The ALP-photon conversion would occur in the presence of an external magnetic field. In order to describe the propagating system along the direction of $\boldsymbol{x_3} = \boldsymbol{x_1} \times \boldsymbol{x_2}$, we utilize a state vector denoted as $\boldsymbol{\Psi} = (A_1, A_2, a)^T$, where $A_1$ and $A_2$ represent the two linear photon polarization amplitudes along $\boldsymbol{x_1}$ and $\boldsymbol{x_2}$, respectively.

The polarization state of the ALP-photon system can be characterized  by the density matrix $\rho \equiv \Psi \otimes \Psi^\dagger$. When the system traverses through a homogeneous magnetic field, $\rho$ obeys the von Neumann-like commutator equation \cite{DeAngelis:2007dqd,Mirizzi:2009aj}: 
\begin{equation}\label{equ:von Neumannn-like}
    i\frac{d\rho}{d\boldsymbol{x_3}} = [\rho, \mathcal{M}_0], 
\end{equation}
where $\mathcal{M}_0$ is the mixing matrix including the interactions between photons and external magnetic field. If the transverse magnetic field $B_t$ aligns with the direction $\boldsymbol{x_2}$, $\mathcal{M}_0$ can be written as
\begin{equation}
    \mathcal{M}_0 = \begin{bmatrix}
\Delta_{\perp} & 0 & 0 \\
0 & \Delta_{\parallel} & \Delta_{a\gamma} \\
0 & \Delta_{a\gamma}  & \Delta_{a} 
\end{bmatrix},
\end{equation}
with $\Delta_{\perp} = \Delta_{pl} +2\Delta_{QED}$, $\Delta_{\parallel} = \Delta_{pl} + 7/2\Delta_{QED}$, $\Delta_{a}=-m_a^2/(2E)$, and $\Delta_{a\gamma}=g_{a\gamma}B_t/2$. The diagonal element $\Delta_{pl} = -\omega_{pl}/(2E)$ describes the photon propagation effect in the plasma with the typical frequency $\omega_{pl}$. $\Delta_{QED}=\alpha E/(45\pi)(B_{\perp}/B_{cr})^2$ is the QED vacuum polarization term with $\alpha$ being the fine structure constant and $B_{cr}=m_e^2/|e|$ being the critical magnetic field. The off-diagonal element $\Delta_{a\gamma}=g_{a\gamma}B_t/2$ represents the ALP-photon mixing effect.

The solution to Eq.~\ref{equ:von Neumannn-like} can be expressed as $\rho(\boldsymbol{x_3}) = \mathcal{T}(\boldsymbol{x_3}) \rho(0) \mathcal{T}(\boldsymbol{x_3})^\dagger$, where $\rho_0$ represents the initial state of the ALP-photon system, and $\mathcal{T}(\boldsymbol{x_3})$ is the transfer matrix that depends on the mixing matrix and the actual angle between $B_\perp$ and $\boldsymbol{x_2}$. 

High energy photons emitted from extragalactic sources  traverse a series of astrophysical magnetic fields before reaching the Earth. The entire path can be divided into many pieces, with the magnetic field in each piece considered to be constant. The survival probability of the photon is written as \cite{Raffelt:1987im,Mirizzi:2007hr}: \begin{equation}\label{equ:Pgaga}
P_{\gamma\gamma} = \mathrm{Tr}\left((\rho_{11}+\rho_{22})\mathcal{T}(\boldsymbol{x_3}) \rho(0) \mathcal{T}^{\dagger}(\boldsymbol{x_3})\right), 
\end{equation}
where $\mathcal{T}(\boldsymbol{x_3}) = \prod \limits_i^n \mathcal{T}_i (\boldsymbol{x_3})$ is the total transferring matrix, and $\mathcal{T}_i(\boldsymbol{x_3})$ is derived from the $i$-th piece along the path. As the polarization of very high energy $\gamma$-rays cannot be measured, the $\gamma$-ray photons emitted from the source are assumed to be unpolarized and $\rho(0)$ is taken to be $\text{diag}(1/2,1/2,0)$ here.

\subsection{Astrophysical environments}

In this study, we focus on analyzing the $\gamma$-ray photons originating from the BL Lac type object Mrk 421. These photons traverse  various astrophysical environments before reaching the Earth, including the broad line region, blazar jet, host galaxy, galaxy cluster, extragalactic space, and Milky Way. 
Our analysis specifically considers the ALP-photon oscillation effect in the blazar and the Milky Way. The previous research has suggested that  internal $\gamma$-ray absorption and ALP-photon oscillation within the broad line region can be disregarded for BL Lac type objects \cite{Tavecchio:2014yoa}. Although BL Lac objects are situated in elliptical galaxies with magnetic fields on the order of $\sim \mu $G, the impact of ALP-photon oscillation can still be neglected \cite{Meyer:2014epa}. Additionally, there is no evidence to suggest that Mrk 421 is located in a rich galaxy cluster with a significant inter-cluster magnetic field, so we do not consider the ALP-photon oscillation effect in this system. Finally, as the magnetic field in the extragalactic space is not well-determined and has only an upper limit on the order of $10^{-9} $ G \cite{Kronberg:1993vk,Blasi:1999hu,Durrer:2013pga}, the ALP-photon oscillation effect in this region is not taken into account in this study. 

Below, we briefly describe the magnetic fields in the blazar jet and Milky Way, and the EBL effect in the extragalactic space. The jet magnetic field of the BL Lac object includes both poloidal and toroidal components. At larger distances from the central black hole, the toroidal component dominates \cite{Pudritz:2012xj, Begelman:1984mw}, while the poloidal component can be safely neglected. The distribution of the toroidal magnetic field and electron density in the co-moving frame can be written as \cite{Pudritz:2012xj, Begelman:1984mw, OSullivan:2009dsx}: 
\begin{equation}
    B^{{\rm jet}}(r) = B_0^{\rm jet} \left( \frac{r}{r_{\rm VHE}}\right)^{-1},
\label{BJMF}
\end{equation}
\begin{equation}
    n^{\rm jet}_{\rm el}(r) = n_0^{\rm jet} \left( \frac{r}{r_{\rm VHE}}\right)^{-2} ,
\end{equation}where $r_{\rm VHE}$ represents the distance between the photon emission site and central black hole, and $B_0^{\rm jet}$ and $n_0^{\rm jet}$ are the magnetic field and electron density at the emission site, respectively. For Mrk 421, we set $B_0^{\rm jet}=0.1$ $\mathrm{G}$, $n_0^{\rm jet}=3000$ $\mathrm{cm^{-3}}$, and $r_{\rm VHE}=10^{17}$ $\mathrm{cm}$ \cite{ARGO-YBJ:2015qiq,Celotti:2007rb}. We conservatively assume that there is no magnetic field beyond the maximum radius of the jet magnetic field with $r_{\rm max}=1 \mathrm{kpc}$. The relation between photon energies in the laboratory frame $E_L$ and the co-moving frame $E_j$ is given by $E_j = E_L/\delta_D$, where $\delta_D$ is the Doppler factor and is set to be $25$ for Mrk 421.

After leaving the extragalactic source, the high energy photons interact with the EBL photons through the pair
production, $\gamma + \gamma_{\rm EBL} \to e^+ + e^-$, resulting in a reduction of the observed flux at high energies. If only the EBL absorption effect is considered, the survival probability of the photon after propagation is given by $P_{\gamma\gamma}=e^{-\tau_{\gamma}}$, where $\tau_\gamma$ is the optical depth. In this study, we adopt the EBL model provided in Ref.~ \cite{Franceschini:2008tp}. 
The ALP-photon oscillation would reduce this EBL absorption effect and affect the final photon spectra. 

The magnetic field in the Milky Way is composed of a regular component and a small turbulent component. Because of the small coherence length, the turbulent component can be safely ignored in this study. Therefore, only the regular component is considered here. The model for this component used in this study is taken from Ref.~\cite{Jansson:2012rt}.

The photon spectrum observed on the Earth is given by
\begin{equation}
\frac{d\Phi}{dE} = P_{\gamma\gamma}\frac{d\Phi_{\rm int}}{dE}, 
\end{equation}
where $P_{\gamma\gamma}$ is the survival probability of photons including the EBL absorption effect and ALP-photon oscillation in propagation, and $d\Phi_{\rm int}/dE$ is the intrinsic spectrum of the source. The final transfer matrix $\mathcal{T}$ in 
Eq. \ref{equ:Pgaga} is expressed as $\mathcal{T} = \mathcal{T}_{\rm MW}\mathcal{T}_{\rm EBL}\mathcal{T}_{\rm jet}$, where $\mathcal{T}_{\rm MW}$, $\mathcal{T}_{\rm EBL}$, and $\mathcal{T}_{\rm jet}$ are the transfer functions
in the Milky Way magnetic field, EBL, and blazar jet magnetic field, respectively. In this study, we use the package gammaALPs \cite{Meyer:2021pbp} to calculate $P_{\gamma\gamma}$.

\section{method}\label{sec:method}

In this section, we describe the procedure for fitting the spectrum and the statistical method used to derive constraints on the ALP parameters. We utilize the very high energy $\gamma$-ray spectra of Mrk 421 observed by MAGIC and Fermi-LAT, which include a total of 15 observation periods \cite{MAGIC:2019ozu}.

\subsection{Spectrum fitting}

For the intrinsic blazar spectra, we consider four commonly used forms: power law with exponential cut-off (EPWL), power law with sub/super-exponential cut-off (SEPWL), log parabola (LP), and log parabola with exponential cut-off (ELP). These spectra contain three or four free parameters. The corresponding expressions are given as follows:
\begin{itemize}
    \item EPWL:  $F_0(E/E_0)^{-\Gamma} e^{-E/E_c}$
     \item SEPWL:  $F_0(E/E_0)^{-\Gamma} e^{(-E/E_c)^d}$
     \item LP:  $F_0(E/E_0)^{-\Gamma - b\mathrm{log} (E/E_0)} $
     \item ELP:  $F_0(E/E_0)^{-\Gamma - b \mathrm{log} (E/E_0)} e^{-E/E_c}$
\end{itemize}
where $F_0$, $\Gamma$, $E_c$, $b$, and $d$ are free parameters, and $E_0$ is taken to be 200 GeV. For each period, we perform fits with the four intrinsic spectrum forms and choose the spectrum form using the Akaike information criterion \cite{AIC}. 

Taking into account the energy resolution of experiments, the expected flux in an energy bin between $E_1$ and $E_2$ is given by 
\begin{equation}
\frac{d\Phi}{dE} = \frac{\int_{E_1}^{E2} dE\int_0^{\infty} S(E', E) \frac{d\Phi}{dE'} dE'}{E_2 - E_1}, 
\end{equation}
where $E'$ and $E$ are the true energy and reconstruction energy of the photon, and $S(E', E)$ denotes the energy smearing Gaussian function. The energy resolutions of MAGIC and Fermi-LAT are adopted as 16\% \cite{MAGIC:2014zas}  and 15\% \cite{Fermi_energy_reso2}, respectively.

We consider the $\chi^2$-function in the fitting for the observations of MAGIC and Fermi-LAT following Ref.~\cite{MAGIC:2019ozu}. 
In this reference, the MAGIC collaboration analyzed the Fermi-LAT results in the same operation periods and acquired the Fermi-LAT data in the form of spectral bowties rather than individual data points. In this case, the $\chi^2$ in the fitting procedure is predominantly influenced by the MAGIC data. The corresponding $\chi^2$ is constructed as \begin{align}
        \chi^2(F_0, \Gamma, ...; m_a, g_{a\gamma}) &= \sum\limits _{i=1}^{N} \frac{\left(\frac{d\Phi}{dE}|_i - \frac{d\Phi}{dE} |_{{\rm obs},i}\right)^2}{\delta _i^2}  + \frac{(\Gamma_{\rm fit} - \Gamma_{\rm LAT})^2}{\Delta \Gamma_{\rm LAT}^2} \notag \\
        & +  \frac{\left(\frac{d\Phi}{dE}|_{\rm LAT} - \frac{d\Phi}{dE} |_{\rm obs,LAT}\right)^2}{\delta _{\rm LAT}^2},
\end{align}
where $N$ represents the number of energy bins in MAGIC observations, $i$ denotes the $i$-th energy bin in MAGIC observations, and the subscript LAT stands for the Fermi-LAT observation at the decorrelation energy of the Fermi-LAT spectrum. $\frac{d\Phi}{dE}$, $\frac{d\Phi}{dE} |_{\rm obs}$, and $\delta $ stand for the predicted value, observed value, and experimental uncertainty of the photon flux, respectively. $\Gamma_{\rm fit}$, $\Gamma_{\rm LAT}$, and $\Delta \Gamma_{\rm LAT}$  are the predicted value, observed value, and experimental uncertainty of the spectral index at the decorrelation energy for the Fermi-LAT observation, respectively.

\subsection{Statistic method}

Given a data sample and a parameter point $(m_a, g_{a\gamma})$, we define the test statistic (TS) value as the logarithm of the likelihood ratio
\begin{equation}\label{equ:TS_CLs}
    {\rm TS}(m_a, g_{a\gamma}) = -2\ln\left( \frac{\mathcal{L}_1 (\doublehat{F_0}, \doublehat{\Gamma}, ...; m_a, g_{a\gamma})}{\mathcal{L}_0(\hat{F_0}, \hat{\Gamma}, ...)} \right),
\end{equation}
where $\mathcal{L}_0$ is likelihood under the null hypothesis, $(\hat{F_0}, \hat{\Gamma}, ...)$ are the intrinsic spectrum parameters maximizing $\mathcal{L}_0$,  $\mathcal{L}_1$ is the likelihood function under the alternative hypothesis, and $(\doublehat{F_0}, \doublehat{\Gamma}, ...)$ are the intrinsic spectrum parameters maximizing $\mathcal{L}_1$ for the given $(m_a, g_{a\gamma})$. The likelihood function is given by
\begin{equation}\label{eq: likelihood}
    \mathcal{L} (F_0, \Gamma, ...; m_a, g_{a\gamma}) = \exp(-\chi^2(F_0, \Gamma, ...; m_a, g_{a\gamma})/2).
\end{equation}
As the impact of the ALP  parameters on the spectrum is non-linear, and the commonly used Wilks' theorem \cite{wilk} is not appropriate in this scenario \cite{Fermi-LAT:2016nkz}. Therefore, the TS value would not obey a $\chi^2$ distribution, and Monte Carlo simulations are needed to derive the realistic TS distribution, which is necessary for setting constraints on the ALP parameters. 

Below, we outline the procedure for establishing excluded regions in the ALP parameter space using the ${\rm CL_s}$ method \cite{Junk:1999kv,Read:2002hq_cls,Lista:2016chp}. We systematically scan the $m_a-g_{a\gamma}$ plane and assess whether each parameter point has been excluded by the observations.

For example, we select a parameter point in the $m_{a}- g_{a\gamma}$ plane.  For this parameter point, with the actual observed data, the best-fit spectrum is obtained by maximizing the likelihood function in Eq.~\ref{eq: likelihood}. Using this best-fit spectrum, we calculate the predicted photon flux in an energy bin. Subsequently, we randomly generate the photon flux for mock data with a Gaussian distribution, where the mean value and deviation are taken to be the predicted flux and the actual uncertainty, respectively. This procedure is repeated for all energy bins to derive a mock data sample, resulting in a TS value for this sample through Eq.~\ref{equ:TS_CLs}.
We generate 1000 mock data samples, and derive the TS distribution from all the samples, denoted as $\rm \{TS\}_{s+b}$. Similarly, we obtain the TS distribution $\rm \{TS\}_{b}$ based on another 1000 mock data samples, which are generated with the best-fit photon spectrum without ALPs.

%We calculate the TS value $\rm TS_{obs}$ for the observed flux points. Finally, we obtain $\rm TS_{obs}$, $\rm \{TS\}_{s+b}$ and $\rm \{TS\}_{b}$.}
%Finally, we obtain two TS distributions with two mock data sets, generated with the predicted spectrum with and without ALPs $(m_{a,i}, g_{a\gamma,i})$.

The $\rm CL_s$ method uses the $\rm CL_s$ value as a criterion to determine whether the selected parameter point is excluded. Given the TS value obtained from the actual data, denoted as $\rm TS_{obs}$, ${\rm CL_s}$ is defined as \cite{Junk:1999kv,Read:2002hq_cls}
\begin{equation}
    \rm CL_s = \frac{CL_{s+b}}{CL_b},
\end{equation}
where ${\rm CL_{s+b}}$ and ${\rm CL_{b}}$ denote the probabilities of obtaining the TS values larger than $\rm TS_{obs}$  according to $\rm \{TS\}_{s+b}$ and $\rm \{TS\}_{b}$, respectively. When the ${\rm CL_s}$ value is smaller than a number $\alpha$, the selected parameter point is considered to be not compatible with the observation at the $1-\alpha$ confidence level (C.L.). For instance, we can take $\alpha=0.05$ to determine whether the the selected ALP parameter is excluded at the $95\%$ C.L..

\begin{figure*}
  \centering
  \subfloat[$m_a$ = $10^{-10}$ eV, $g_{a\gamma}$ = $5\times10^{-10}$ GeV.]
  {\includegraphics[width=0.4\textwidth]{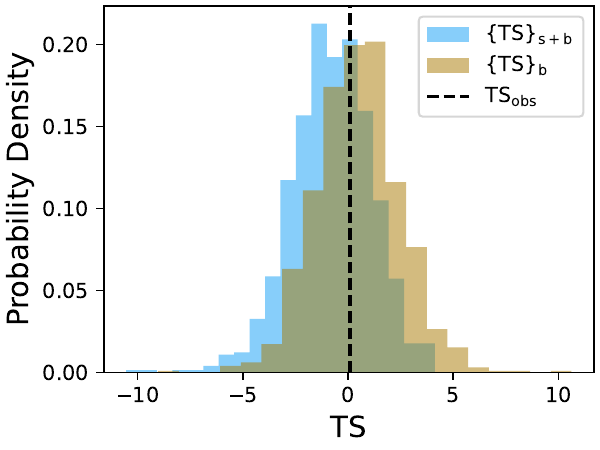}\label{fig: TS distri cls27}}
  \subfloat[$m_a$ = $2\times10^{-9}$ eV, $g_{a\gamma}$ = $3.2\times10^{-11}$ GeV.]
  {\includegraphics[width=0.4\textwidth]{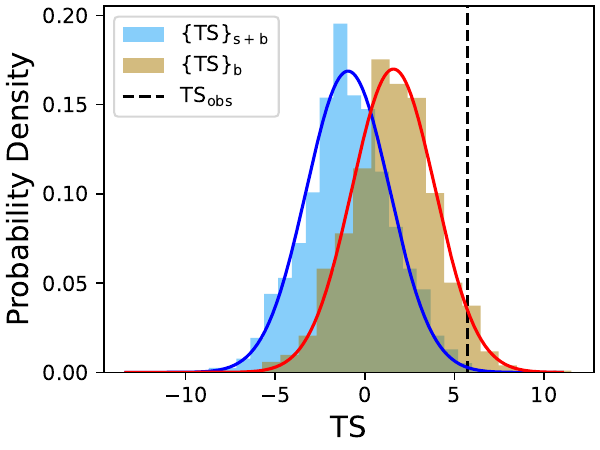}\label{fig: TS distri cls418}} 
   \quad
  \subfloat[$m_a$ = $10^{-8}$ eV, $g_{a\gamma}$ = $1.3\times10^{-10}$ GeV.]
  {\includegraphics[width=0.4\textwidth]{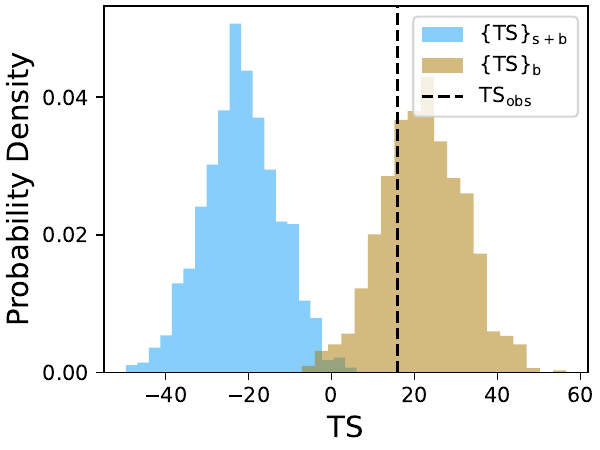}\label{fig: TS distri cls641}} 
  \subfloat[TS distribution of simplified method.]{\includegraphics[width=0.4\textwidth]{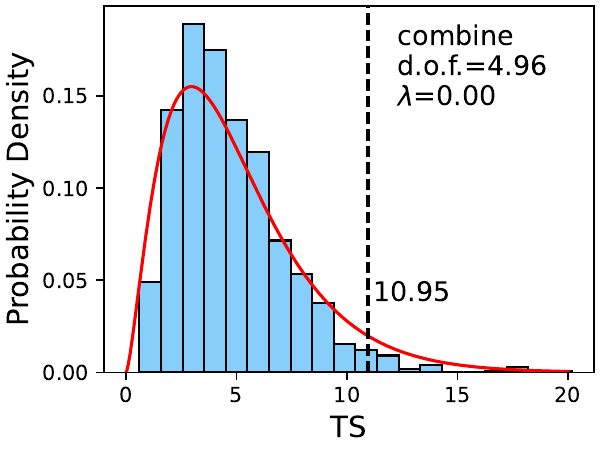}\label{fig: TS distri old}}
  \caption{The TS distributions $\rm \{TS\}_{s+b}$ and $\rm \{TS\}_{b}$ for three parameter points are shown in panels (a), (b), and (c). In these panels, the vertical dashed lines represent the observed TS. The TS distribution under the null hypothesis derived with the simplified method is shown in panel (d). In this panel, the vertical dashed lines represent the threshold TS value at the $95\%$ C.L.. The solid lines in panels (b) and (d) represent the TS distributions derived from fitting.}
  \label{fig: TS distri}
\end{figure*}

The TS distributions for three typical parameter points derived from the joint analysis of all 15 periods are shown in Fig. \ref{fig: TS distri cls27},  \ref{fig: TS distri cls418}, and  \ref{fig: TS distri cls641}, where the corresponding  ${\rm CL_s}$ values are 0.574, 0.056, and 0.0, respectively. These results indicate that the third point can be excluded by the observations at the $95\%$ C.L., while the first two points are not excluded at the $95\%$ C.L..

For the traditional frequentist analysis, the constraint at the $1-\alpha$ C.L. is obtained by requiring $\rm CL_{s+b}<\alpha$ based on $\rm \{TS\}_{s+b}$. Therefore, the same confidence levels of the constraints derived from the $\rm CL_s$ method and traditional frequentist methods have different statistical meanings. If the TS distributions $\rm \{TS\}_{s+b}$ and $\rm \{TS\}_{b}$ are well separated and $\rm TS_{obs}$ is near to the left tail of the TS distribution $\rm \{TS\}_{b}$ (see e.g. Fig. \ref{fig: TS distri cls641}), the constraints derived from the two methods would be approximately equal. However, in some parameter regions, the two TS distributions have large overlaps (see e.g. Fig. \ref{fig: TS distri cls27} and \ref{fig: TS distri cls418}). This indicates that the experiment is not sensitive to the signal and the two hypotheses are difficult to be distinguished. In this case, $\rm CL_s$ is generally larger than $\rm CL_{s+b}$ due to the small $\rm CL_{b}$. A parameter point excluded by the $\rm CL_{s+b}< \alpha$ criterion may be still allowed by the $\rm CL_s$ criterion with $\rm CL_{s}> \alpha$. Therefore, the $\rm CL_s$ method sets a less radical constraint compared with the traditional frequentist constraint when the experiment is difficult to distinguish the null and alternative hypotheses.

Due to the large parameter space in the scan and high computational resource consumption, the size of our simulation sample for each parameter point is limited to be 1000. This would lead to some statistical errors in the tail of the TS distribution due to the fluctuation. Such statistical errors may significantly affect the precision of the calculation of $\rm CL_s$ when $\rm CL_s$ is very small, since the $\rm TS_{obs}$ would approach the tail of the TS distribution in this case. In order to reduce such uncertainties, we fit the two TS distributions using the Gaussian function when ${\rm CL_s}$ is near to 0.05 and derive $\rm CL_s$ from these fitting distributions (see e.g. Fig. \ref{fig: TS distri cls418}).

For each parameter point in the $m_a-g_{a\gamma}$ plane, we calculate the $\rm CL_s$ value and evaluate whether it has been excluded at the $95\%$ C.L. following the above procedure. Subsequently, the excluded regions consisting of the parameter points with $\rm CL_s<0.05$ can be directly established.

For comparison, we also calculate the result with the simplified method.
In this method, the TS distribution is derived from the mock data samples generated with the best-fit photon spectrum without ALPs, and is assumed to be valid under the alternative hypothesis with ALPs. For each mock data sample, TS is defined as
\begin{equation}\label{equ:TS_sim}
    {\rm TS} = 2\ln\left( \frac{\mathcal{L}_1 (\doublehat{F_0}, \doublehat{\Gamma}, ...; \hat{m}_a, \hat{g}_{a\gamma})}{\mathcal{L}_0(\hat{F_0}, \hat{\Gamma}, ...)} \right),
\end{equation}
where ($\hat{m}_a, \hat{g}_{a\gamma}$) maximize $\mathcal{L}_1$ in the whole $m_a-g_{a\gamma}$ plane. Subsequently, one can fit the TS distribution with a non-central $\chi^2$ distribution, and determine the threshold value $\Delta \rm TS_{\alpha'}$,  which is the quantile of $\alpha' \%$ in the TS distribution.
Finally, the excluded regions at the $\alpha' \%$ C.L. are established by determining the parameter points for the actual data with $ \hat{\chi}^2_{\rm ALP} > \hat{\chi}^2_{\alpha'} \equiv \hat{\chi}^2_{\rm ALP, min} + \Delta \rm TS_{\alpha'}$, where $\hat{\chi}^2_{\rm ALP,min}$ denotes the global best-fit $\chi^2$ value for the actual data in the $m_a-g_{a\gamma}$ plane. For instance, the ALP parameter points yielding $\hat{\chi}^2_{\rm ALP} >  \hat{\chi}^2_{95}$ is considered to be excluded at the 95\% C.L.. In Fig. \ref{fig: TS distri old}, we show the TS distribution and corresponding $\Delta \rm TS_{95}$ derived from the observations of 15 periods. This TS distribution can be fitted with a non-central $\chi^2$ function with a degree of freedom of 4.95 and a non-centrality parameter $\lambda$ of 0.0. The corresponding $\Delta \rm TS_{95}$ is 10.95.

\section{results}\label{sec:result}

\begin{table}[htbp]
  \centering
  \caption{The intrinsic spectrum forms, the best-fit $\chi^2$ values under the null and alternative hypotheses, and the threshold TS values corresponding to $95\%$ C.L. derived with the simplify method for 15 observation periods are listed here. The last row shows the results of the joint analysis.}
  \label{tab:SED}
  \begin{tabular}{cccccc}
    \hline
    \hline
    period & SED & $\chi^2_{\rm Null}$ & $\chi^2_{\rm ALP,min}$ & $\Delta TS_{95}$  \\ 
    \hline
    20130410 & SEPWL & 11.80 & 9.41 & 9.35 \\
    20130411 & ELP & 16.40 & 10.51 & 10.58 \\
    20130412 & ELP & 9.53 & 7.00 & 11.01 \\
    20130413a & ELP & 16.25 & 13.87 & 11.06 \\
    20130413b & SEPWL & 10.27 & 8.32 & 10.78 \\
    20130413c & SEPWL & 11.55 & 6.99 & 10.00 \\
    20130414 & ELP & 23.04 & 14.00 & 10.88 \\
    20130415a & ELP & 6.89 & 5.56 & 10.60 \\
    20130415b & SEPWL & 13.26 & 9.40 & 11.19 \\
    20130415c & SEPWL & 4.98 & 3.77 & 10.88 \\
    20130416 &  SEPWL & 33.24 & 22.51 & 10.84 \\
    20130417 & SEPWL & 27.85 & 16.15 & 11.01 \\
    20130418 & SEPWL & 9.38 & 7.70 & 9.64 \\
    20130419 & EPWL & 4.18  & 2.31 & 7.96 \\
    20140426 & ELP & 24.07 & 14.24 & 10.41 \\
    Joint & - & 222.71 & 216.50 & 10.95 \\
    \hline
    \hline
  \end{tabular}
\end{table}

\begin{figure*}
  \centering
  \subfloat{\includegraphics[width=0.45\textwidth]{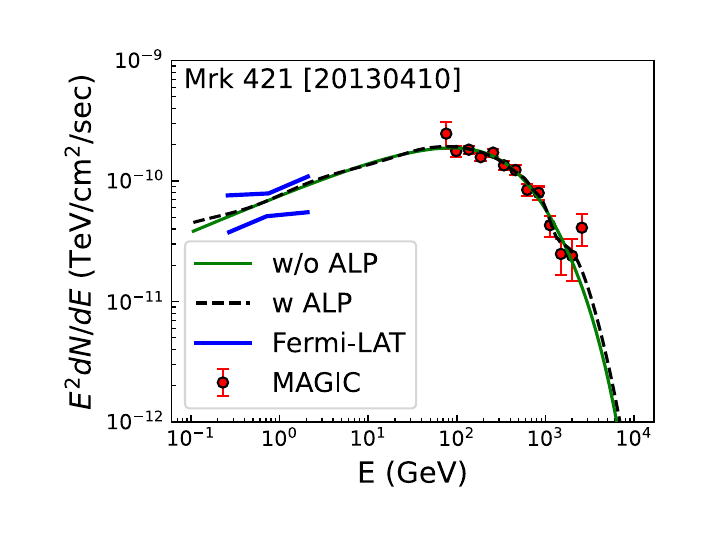}\label{fig: spec20130410}}
  \subfloat{\includegraphics[width=0.45\textwidth]{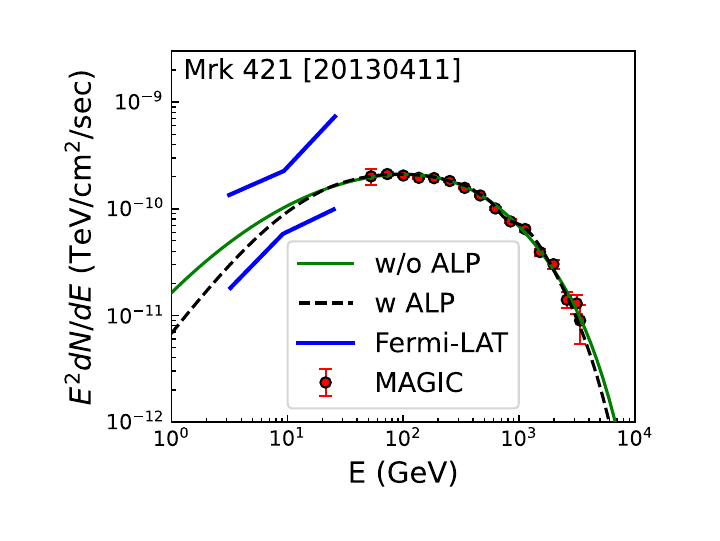}\label{fig: spec20130411}} 
   \quad
  \subfloat
  {\includegraphics[width=0.45\textwidth]{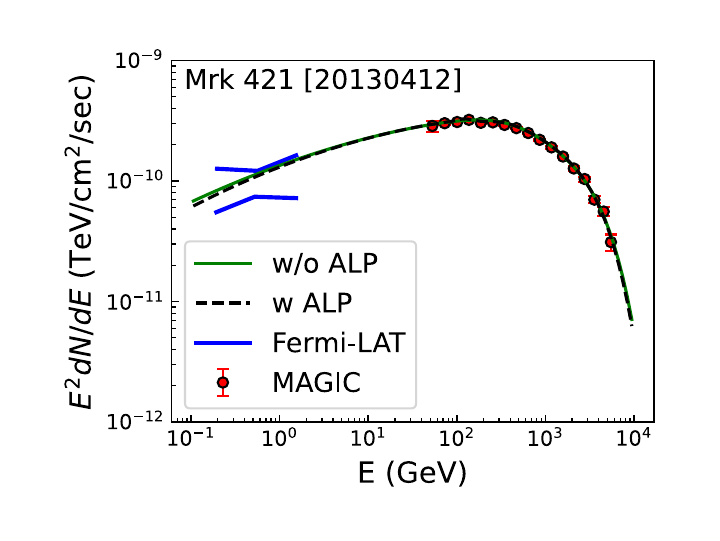}\label{fig: 20130412}} 
  \subfloat{\includegraphics[width=0.45\textwidth]{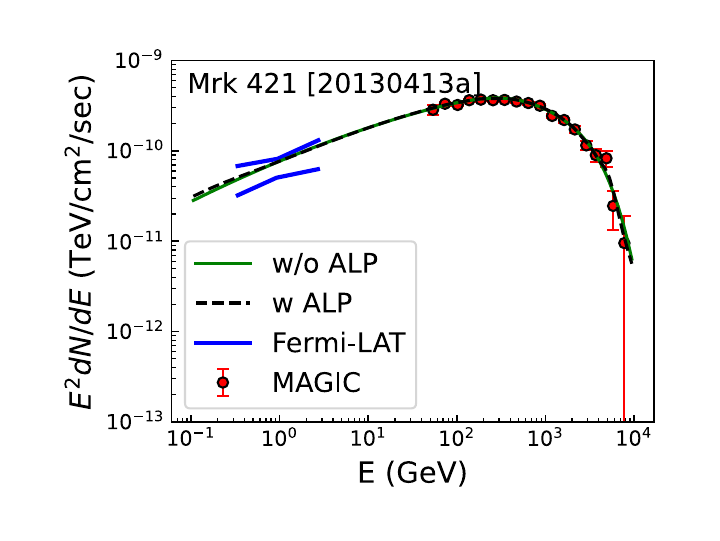}\label{fig: 20130413a}}
  \caption{The best-fit $\gamma$-ray spectra of Mrk 421 in the periods 20130410, 20130411, 20130412, and 20130413a. The solid and dashed lines represent the spectra under the null and alternative hypotheses, respectively. The red points and blue bow-ties represent the results of MAGIC and Fermi-LAT, respectively \cite{MAGIC:2019ozu}.}
  \label{fig: spec}
\end{figure*}

In this section, we present the constraints on the ALP parameters using the observations of Mrk 421 by MAGIC and Fermi-LAT. The best-fit $\chi^2$ values under the null and alternative hypotheses for each period are listed in Table. \ref{tab:SED}. 
Our results indicate that the null hypothesis provides a good fit to the data for the majority of observation periods, with a $\chi^2_{\rm Null}$ value per degree of freedom near or below than 1. The inclusion of ALP does not significantly improve the fit in these cases.
These results are illustrated in Fig. \ref{fig: spec} depicting the best-fit spectra under the null and alternative hypotheses for the initial four periods.
However, for the observation periods 20130414, 20130416, 20130417, and 20140426, the null hypothesis does not fit the data very well, yielding a $\chi^2_{\rm Null}$ value per degree of freedom about 2. In these cases, the inclusion of ALP slightly improve the fit, although the significance is not very high.

\begin{figure*}[htbp]
  \centering
  \begin{minipage}[b]{0.22\textwidth}
    \centering
    \includegraphics[width=\textwidth]{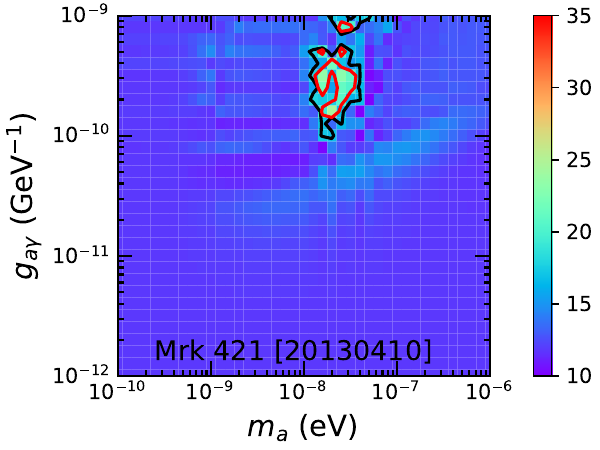}
  \end{minipage}
  \begin{minipage}[b]{0.22\textwidth}
    \centering
    \includegraphics[width=\textwidth]{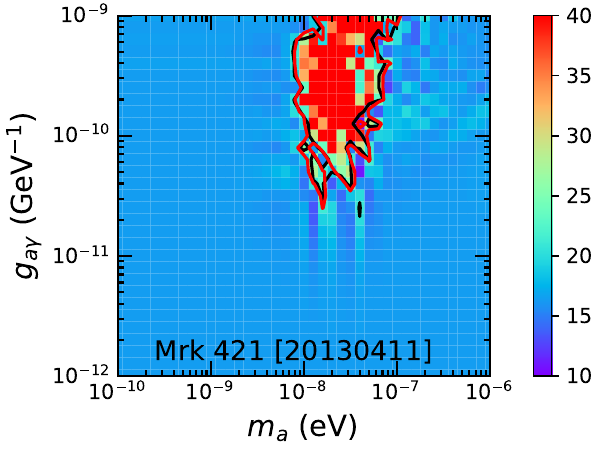}
  \end{minipage}
  \begin{minipage}[b]{0.22\textwidth}
    \centering
    \includegraphics[width=\textwidth]{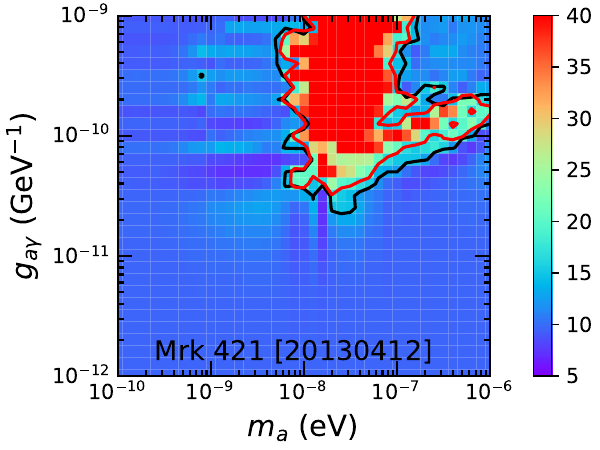}
  \end{minipage}
  \begin{minipage}[b]{0.22\textwidth}
    \centering
    \includegraphics[width=\textwidth]{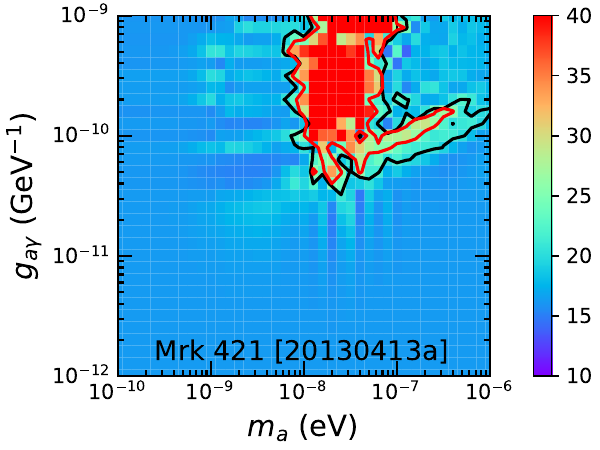}
  \end{minipage}\\
  \begin{minipage}[b]{0.22\textwidth}
    \centering
    \includegraphics[width=\textwidth]{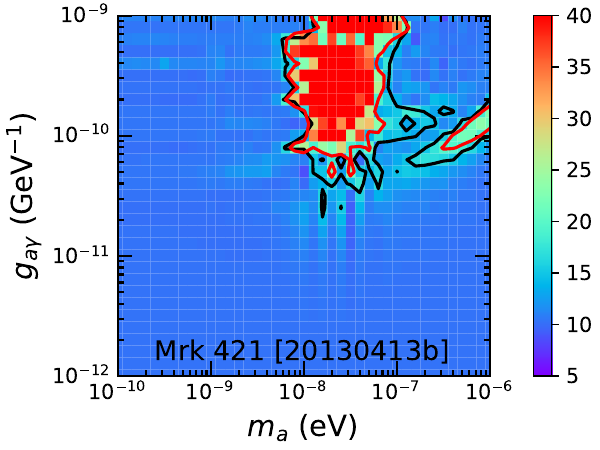}
  \end{minipage}
  \begin{minipage}[b]{0.22\textwidth}
    \centering
    \includegraphics[width=\textwidth]{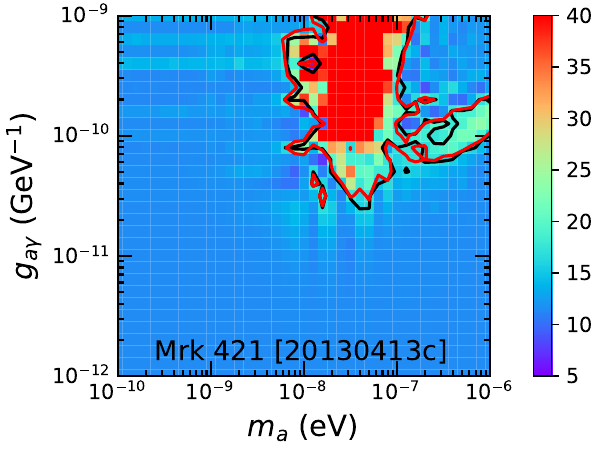}
  \end{minipage}
  \begin{minipage}[b]{0.22\textwidth}
    \centering
    \includegraphics[width=\textwidth]{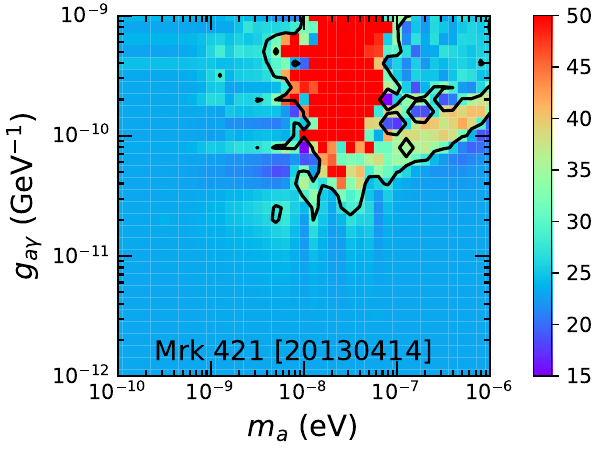}
  \end{minipage}
  \begin{minipage}[b]{0.22\textwidth}
    \centering
    \includegraphics[width=\textwidth]{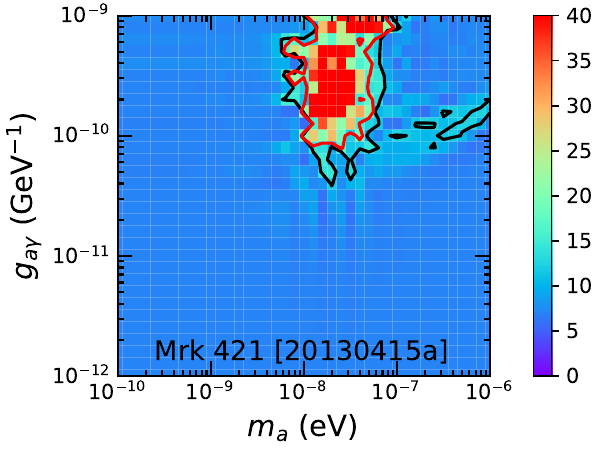}
  \end{minipage}\\
  \begin{minipage}[b]{0.22\textwidth}
    \centering
    \includegraphics[width=\textwidth]{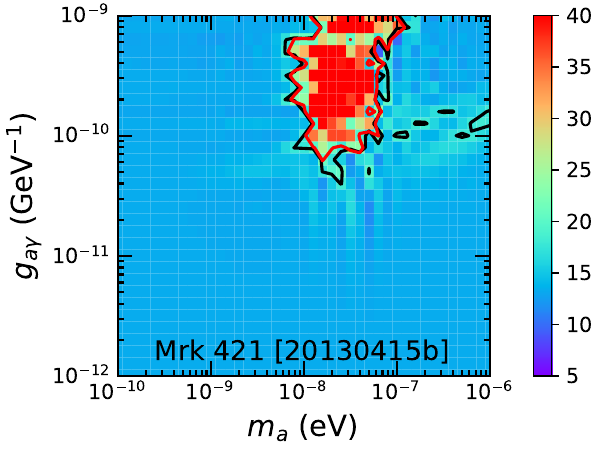}
  \end{minipage}
  \begin{minipage}[b]{0.22\textwidth}
    \centering
    \includegraphics[width=\textwidth]{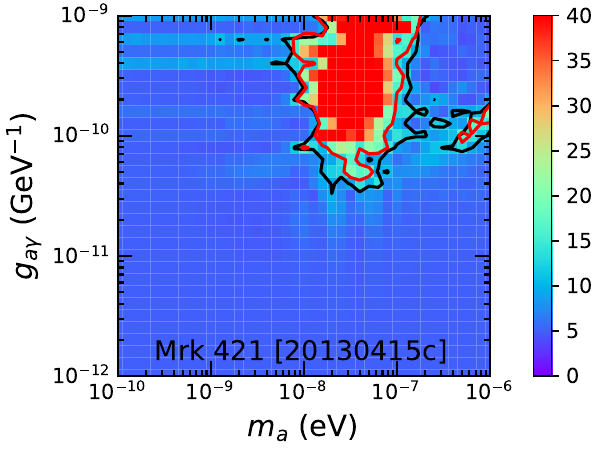}
  \end{minipage}
  \begin{minipage}[b]{0.22\textwidth}
    \centering
    \includegraphics[width=\textwidth]{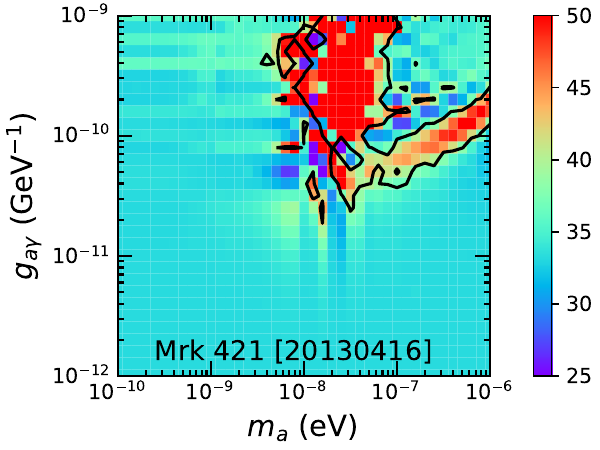}
  \end{minipage}
  \begin{minipage}[b]{0.22\textwidth}
    \centering
    \includegraphics[width=\textwidth]{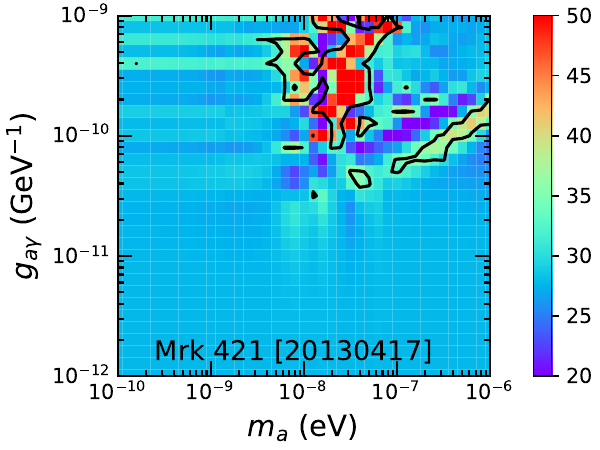}
  \end{minipage}\\
  \begin{minipage}[b]{0.22\textwidth}
    \centering
    \includegraphics[width=\textwidth]{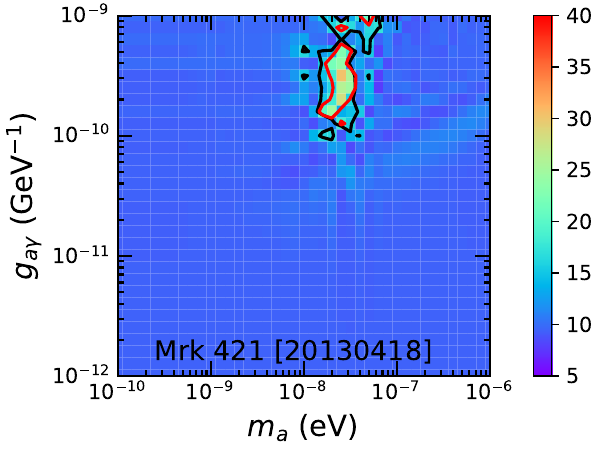}
  \end{minipage}
  \begin{minipage}[b]{0.22\textwidth}
    \centering
    \includegraphics[width=\textwidth]{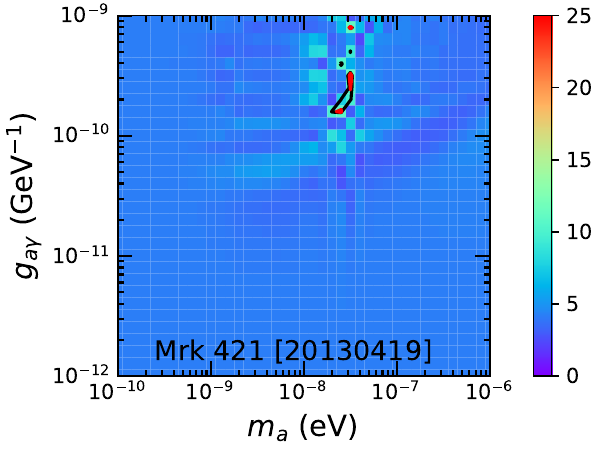}
  \end{minipage}
  \begin{minipage}[b]{0.22\textwidth}
    \centering
    \includegraphics[width=\textwidth]{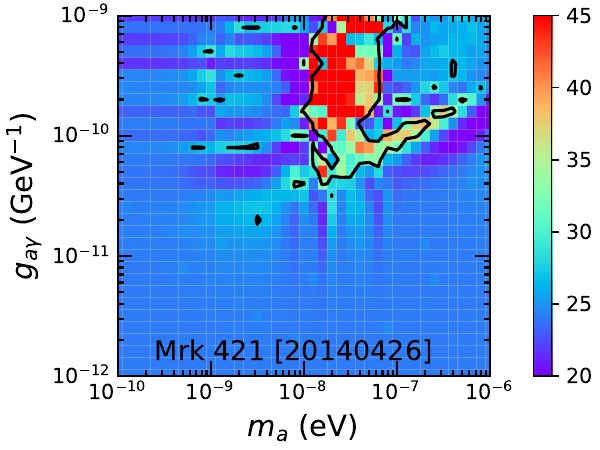}
  \end{minipage}
   \begin{minipage}[b]{0.22\textwidth}
    \centering
    \includegraphics[width=\textwidth]{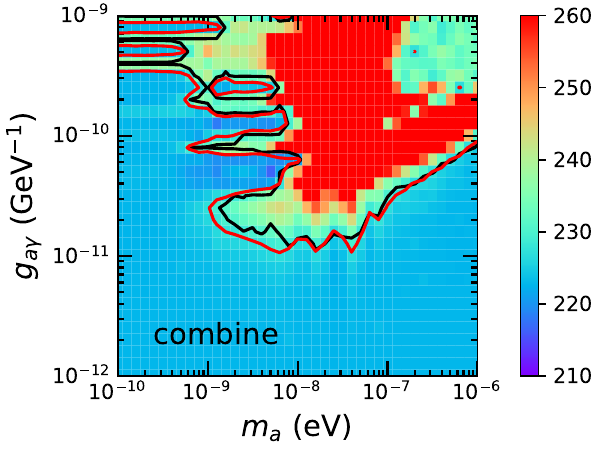}
  \end{minipage}
  \caption{The heat maps of the $\chi^2$ values in the $m_a-g_{a\gamma}$ plane under the alternative hypothesis for the actual data during 15 observation periods.
  The solid black and red lines represent the constraints derived with the ${\rm CL_s}$ and simplified methods, respectively. The results of the joint analysis is also shown in the right bottom panel.}
  \label{fig: contour multi-figures}
\end{figure*}

In our analysis, we conduct a parameter scan within the range of $m_a \sim 10^{-10}-10^{-6}$ eV and $g_{a\gamma} \sim 10^{-12}-10^{-9}$ GeV$^{-1}$, with a bin width of 0.1 in logarithmic space. The heat maps of the $\chi^2$ values under the alternative hypothesis for the actual data during 15 observation periods 
are shown in Fig. \ref{fig: contour multi-figures}. The black lines represent the constraints at the 95\% C.L. obtained using the ${\rm CL_s}$ method. Due to the time variability of the photon flux emitted by Mrk 421, the observations from different periods are assumed to be independent, resulting in varying constraints for different periods. Notably, the constraints obtained from periods 20130410, 20130418, and 20130419 are relatively weak, while those from periods 20130412, 20130414, and 20130416 are  comparatively  stringent. Note that the constraints may depend sensitively on the ALP parameters in some parameter regions, resulting in isolated excluded regions. For instance, in addition to the parameter region between $m_a \sim 10^{-8}-10^{-7}$ eV, there are small regions with $m_a < 10^{-9}$ eV that can be excluded for period 20140426. 

For comparison, the constraints at the 95\% C.L. obtained using the simplified method are also displayed in Fig. \ref{fig: contour multi-figures}.
We observe that for the majority of observation periods, the simplified method provides constraints that are comparable to those obtained with the ${\rm CL_s}$ method. This test confirms the validity of the simplified method in the previous studies. However, it is important to note that  the simplified method is unable to establish constraints for certain periods, specifically 20130414, 20130416, 20130417, and 20140426, as depicted in Fig. \ref{fig: contour multi-figures}. For these periods, the presence of ALPs can improve fits in specific ALP parameter regions, resulting in much smaller $\chi^2$ values compared to those under the null hypothesis. Consequently, the corresponding $\hat{\chi}^2_{\rm 95}$ are close to $\hat{\chi}^2_{\rm Null}$, yielding unreasonable constraints. However, the ${\rm CL_s}$  is capable of providing constraints for all periods, even when the ALP hypothesis improves the fit in the periods 20130414, 20130416, 20130417, and 20140426. This is an advantage of using the ${\rm CL_s}$ method in ALP studies.

\begin{figure}
  \centering
  \includegraphics[width=0.45\textwidth]{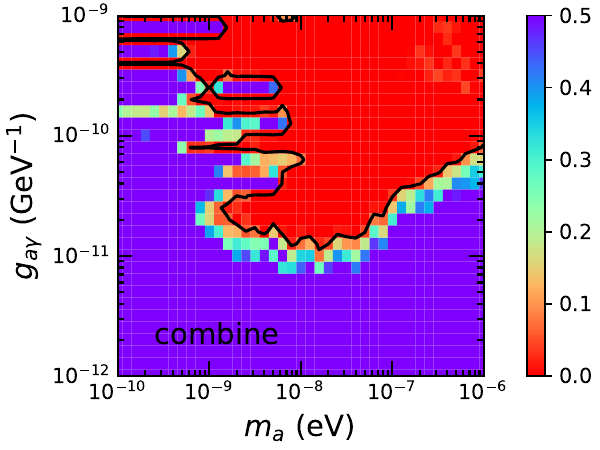}
  \caption{The heat map of the $\rm CL_s$ values in the $m_a-g_{a\gamma}$ plane from the joint analysis. The excluded regions above the black solid line consist of the parameter points excluded at the $95\%$ C.L. with $\rm CL_s< 0.05$.}
  \label{fig:combine_CLs_contour}
\end{figure}

\begin{figure}
  \centering
  \includegraphics[width=0.45\textwidth]{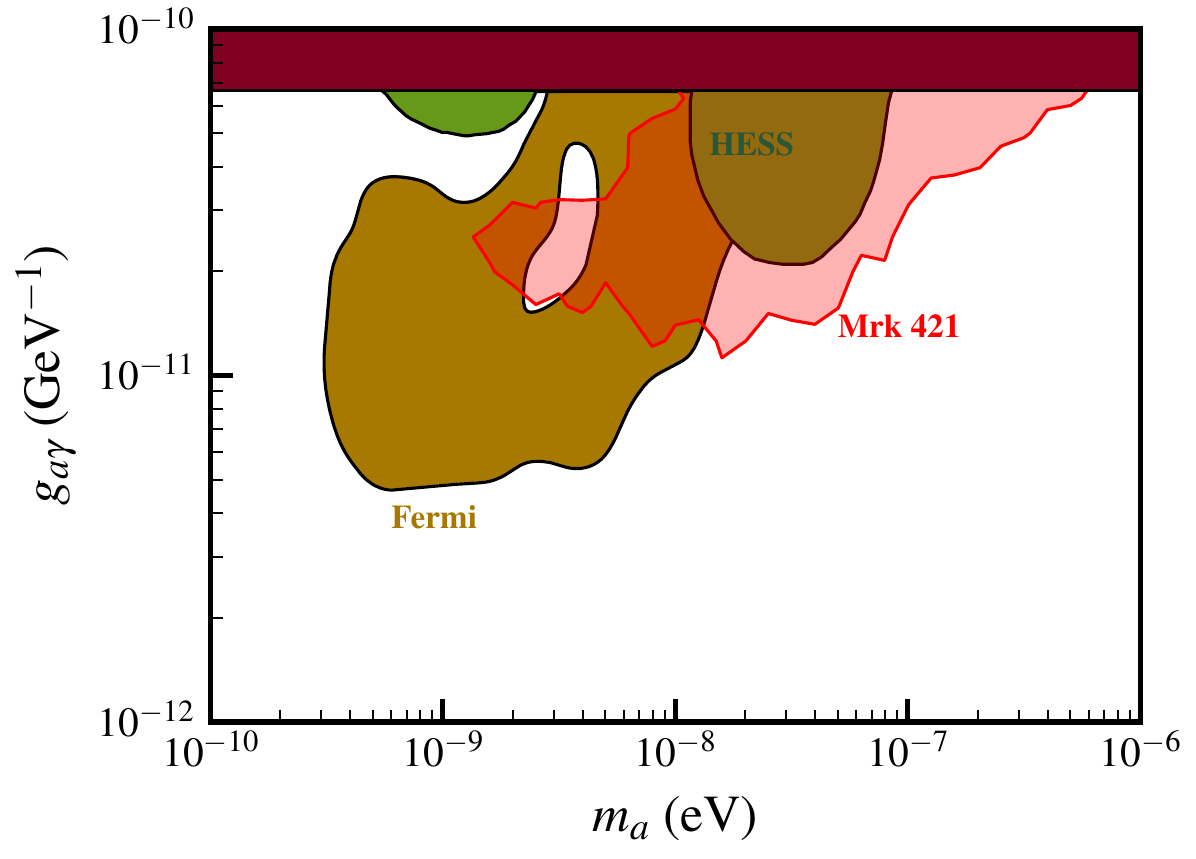}
  \caption{The constraint in the $m_a-g_{a\gamma}$ plane from the Mrk 421 observation of MAGIC and Fermi-LAT, derived with the ${\rm CL_s}$ method at $95\%$ C.L.. For comparison, the constraints set by CAST (dark red contour) \cite{CAST:2017uph}, the PKS 2155-304 observations of H.E.S.S. (green contour) \cite{HESS:2013udx}, and the NGC 1275 observation of Fermi-LAT (orange contour) \cite{Fermi-LAT:2016nkz} are also shown (see also \cite{AxionLimits}).}
  \label{fig:comp}
\end{figure}

As the constraints are complementary in the parameter space for the 15 periods, we derive the constraint from a joint analysis and show the results in Fig. \ref{fig: contour multi-figures} and Fig. \ref{fig:combine_CLs_contour}. The joint $\chi^2$ is calculated as the sum of the $\chi^2$ values from all 15 periods. The corresponding best-fit $\chi^2$ values under the null and alternative hypotheses are listed in Table. \ref{tab:SED}. It can be seen that the joint analysis significantly enhances the constraints. The heat map of the $\rm CL_s$ values from the joint analysis is shown in Fig. \ref{fig:combine_CLs_contour}. The excluded regions above the black solid line consist of the parameter points excluded at the $95\%$ C.L. with $\rm CL_s< 0.05$. In Fig. \ref{fig:comp}, we present the constraints set by the joint analysis alongside those from other experiments, including the CAST experiment (where $g_{a\gamma}>6.6\times10^{-11}$ GeV$^{-1}$ is excluded), Fermi-LAT observation for NGC 1275 \cite{Fermi-LAT:2016nkz}, and H.E.S.S. observation for PKS 2155-304 \cite{HESS:2013udx}. Our results show that the ALP-photon coupling $g_{a\gamma}$ is constrained to be smaller than $ \sim 2\times10^{-11}$ GeV$^{-1}$ for ALP masses between $10^{-9}$ eV and $10^{-7}$ eV. These constraints are complementary to those obtained from other experimental results in certain parameter regions.

Note that in our study, we have assumed that the intrinsic spectra during different variability periods are independent. Consequently, sources with observations of multiple variability periods, which effectively expand the data samples, are suitable for studying ALPs. This is the case for the blazar Mrk 421, which has been observed during multiple variability periods. However, if there are common underlying mechanisms determining the variability of the source, the intrinsic spectra of the source would not be independent.
In the absence of a standard variability model, one can reduce the number of free parameters for the intrinsic spectra to account for such a global impact. 
This may lead to greater difficulty in fitting all the observations under the null hypothesis using fewer free parameters. Consequently, the introduction of ALPs may improve the fit and result in different constraints. %precisely established variability models, 

\begin{figure}
  \centering
  \includegraphics[width=0.45\textwidth]{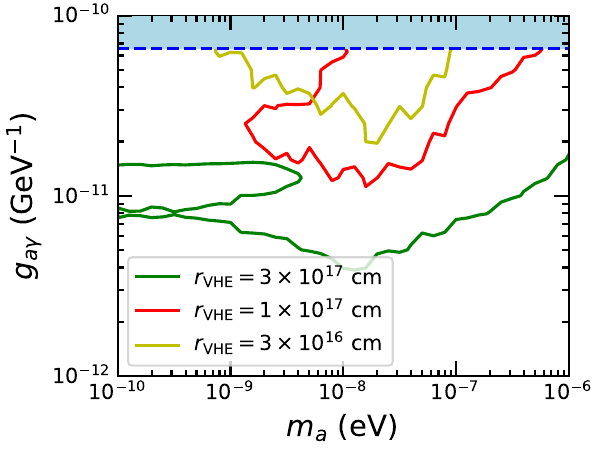}
  \caption{The constraints in the $m_a-g_{a\gamma}$ plane for three values of $r_{\rm VHE}$. The yellow, red, and green solid lines depict the constraints corresponding to $r_{\rm VHE}$ values of $3\times 10^{16}$ cm, $10^{17}$ cm, and $3\times 10^{17}$ cm, respectively.}
  \label{fig:combine_res_B0}
\end{figure}

Finally, we emphasize that the constraints derived in this analysis are subject to astrophysical uncertainties. The dominant uncertainty arise from the blazar jet magnetic field model. The impact of parameters such as $B_0$, $\delta_D$, $n_0$, and $r_{\rm VHE}$ on the constraints has been discussed in Ref. \cite{Li:2020pcn}. It has been shown that while the influence of $\delta_D$ and $n_0$ are relatively minor, the final results are substantially affected by the strength of the blazar jet  magnetic field, which is determined by the two parameters $B_0$ and $r_{\rm VHE}$. 
Specifically, the distance of the emission region from the black hole $r_{\rm VHE}$ cannot be precisely determined. 
In Eq. \ref{BJMF}, it is apparent that these two parameters $B_0$ and $r_{\rm VHE}$ appear as a product of them. Consequently, we can effectively focus on the impact of the variation of $r_{\rm VHE}$. We replicate the full analyses for two additional values, $3\times 10^{17}$ cm and $3\times 10^{16}$ cm of $r_{\rm VHE}$, and present the constraints in Fig. \ref{fig:combine_res_B0}. As shown in Fig. \ref{fig:combine_res_B0}, the constraints become more (less) stringent with increasing (decreasing) $r_{\rm VHE}$.

\section{conclusion}\label{sec:summary}

We investigate the impact of the ALP-photon oscillation effect on the $\gamma$-ray spectra of the blazar Mrk 421, as measured by MAGIC and Fermi-LAT during 15 observation periods. Our analysis takes into account the ALP-photon oscillation in both the blazar jet and Galactic magnetic fields. The results of our analysis indicate that, for the majority of observation periods, the $\gamma$-ray spectra can be well-fitted without ALPs. However, for the periods 20130414, 20130416, 20130417, and 20140426, the inclusion of ALP could improve the fit, albeit not with high significance.

In this work, we use the ${\rm CL_s}$ method to set constraints on the ALP parameters for the first time. Combining the data from all observation periods, we determine that the ALP-photon coupling $g_{a\gamma}$ is constrained to be smaller than $\sim 2\times10^{-11}$ GeV$^{-1}$ for ALP masses ranging from $10^{-9}$ eV to $10^{-7}$ eV at the 95\% C.L.. Our results complement those from other experiments in certain parameter spaces. For comparison, we also apply the simplified method that generates the TS distribution based on the null hypothesis rather than the alternative hypothesis. This method is commonly used in previous studies. Our results demonstrate consistency between the constraints obtained from the ${\rm CL_s}$ method and the simplified method for the majority of periods. However, for certain periods, such as 20130414, 20130416, 20130417, and 20140426, the simplified method fails to provide reasonable constraints, since the presence of the ALP could improve the fit compared to the null hypothesis. In contrast, the ${\rm CL_s}$ method remains effective and is able to provide reasonable constraints in this case. This is an advantage of using the ${\rm CL_s}$ method in ALP studies.

In the future, observations such as LHAASO \cite{LHAASO:2019qtb} and CTA \cite{CTAConsortium:2013ofs} hold great potential for providing more precise measurements of blazars in the domain of very high energy $\gamma$-rays. With improved sensitivity, these experiments will enable more accurate assessments of the ALP-photon oscillation effect and yield more stringent constraints on the ALP parameters.

\acknowledgements
We would like to express our gratitude to Mireia Nievas Rosillo for providing the high-energy $\gamma$-ray spectra data of Mrk 421 measured by MAGIC and Fermi-LAT. We also thank the helpful discussions with Run-Min Yao. X. J. Bi is supported by the National Natural Science Foundation of China under grant No. 12175248. W. B. Lin is supported by the National Natural Science Foundation of China (Grant No. 11973025).

Email: $^*$gaolinqing@hotmail.com 
$^\dag$bixj@ihep.ac.cn
$^\ddag$asd1414987195@gmail.com
$^\S$lwb@usc.edu.cn
$^\P$yinpf@ihep.ac.cn

\newpage
\bibliographystyle{apsrev}
\bibliography{Mrk421_ALP}

\end{document}